\documentclass[%
aps, prb, twocolumn, showpacs, superscriptaddress, byrevtex,
 amsmath,amssymb,
]{revtex4-1}

\usepackage{graphicx}
\usepackage{dcolumn}
\usepackage{bm}

\begin{document}

\preprint{AIP/123-QED}

\title{Evidence of inter-layer interaction in magneto-luminescence spectra of electron bilayers}

\author{Ilirjan Aliaj}

\author{Vittorio Pellegrini}

\author{Andrea Gamucci}
\email{andrea.gamucci@sns.it}

\author{Biswajit Karmakar}

\affiliation{ 
CNR-NANO NEST and Scuola Normale Superiore, 
Piazza San Silvestro 12, 56127 Pisa, Italy
}

\author{Aron Pinczuk}

\affiliation{Dept. of Appl. Phys. \& Appl. Math. and Dept. of Physics, 
Columbia University, New York 10027, USA}

\author{Loren N. Pfeiffer}

\author{Ken W. West}

\affiliation{Dept. of Electrical Engineering,
Princeton University, Princeton, New Jersey 08544, USA}

\date{\today}

\begin{abstract}
Magneto-luminescence studies in electron bilayers reveal the hallmarks of the even-denominator and other quantum Hall states in the intensities and energies of the inter-band optical recombination lines. In the presence of a small tunneling gap between the layers the magneto-optical emission from the lowest anti-symmetric subband, not populated in a single-electron picture, displays maxima at filling factors 1 and 2/3. These findings uncover a loss of pseudospin polarization, where the pseudospin describes the layer index degree of freedom, that is linked to an anomalous population of the anti-symmetric level due to excitonic correlations. The results demonstrate a new realm to probe the impact of inter-layer Coulomb interaction in quantum Hall bilayers. 

\end{abstract}
\maketitle

%

The terms of Coulomb interaction that arise from the spatial separation of electrons
in double layer semiconductor heterostructures are at the origin of several new phenomena 
that occur in the quantum Hall (QH) regime \cite{book}.
The prominent physics linked to the impact of inter-layer electron interaction dramatically manifests in 
the even-denominator QH state at total filling factor $\nu _{T}= 1/2 $ \cite{eisen1}. Much attention was devoted to the quantum phase diagram 
of bilayers at $\nu _{T} = 1$ as a function of $\Delta _{SAS}/E_c$ and $d/l_B$  
($\Delta_{SAS}$ is the tunneling gap, $E_c = e^2/\epsilon l_B$, $l_B$ is the magnetic length, $d$ is the inter-layer distance) \cite{boeb,murphy}.
The description of inter-layer correlated states frequently employs a pseudospin degree of freedom that labels the electron occupation of the 
the left and right layers. In pseudospin language, for example,
the inter-layer correlated QH state at $\nu _T =1$ and $\Delta _{SAS} = 0$ is described as an easy-plane
pseudospin ferromagnetic phase with a spontaneously broken symmetry \cite{kun}. Alternatively, this quantum phase can be regarded as an inter-layer exciton condensate \cite{nat}. 
\par
Several experiments have highlighted the unique properties of the intriguing $\nu _T =1$ state that emerges at $\Delta_{SAS} = 0$. These experiments have uncovered evidence of counterflow superfluid-like behavior \cite{kellogg} and
have established the existence of a finite-temperature phase transition \cite{champ,karmak}. At finite values of the tunneling gap, on the other hand, the pseudospins align along a specific direction in the plane, in a manner that is linked to the electron occupation of the symmetric combination (S) of the lowest-energy quantum-well Landau levels (LL). 
However, if the tunneling gap remains sufficiently small, quantum fluctuations can lead to a suppression of the pseudospin ordering, which in turn
leads to an anomalous occupation of the lowest antisymmetric spin-up (AS$\uparrow$) Landau level. Indeed a loss of pseudospin order was probed at $\nu _T =1$  
by inelastic light scattering methods \cite{Lu05}.
\par
While extensive investigations of quantum Hall bilayers were carried out by magneto-transport techniques, light scattering \cite{karmak,Lu05,biswajit,luos06} and NMR \cite{nmr},
little efforts were devoted to studies of magneto-photoluminescence (magneto-PL) \cite{PLbil}. This is surprising since in single layers the magneto-PL
is a powerful probe of electron-correlation and of spin polarization in the regimes of the integer and fractional quantum Hall effects \cite{heim88,turb90,gold90,kuk97,Byszewski,stern10}. Some impacts of Coulomb interactions in magneto-PL, however, are hidden in symmetric modulation-doped heterostructures where the optical emission lines display a cross-over from Landau-level (linear) to excitonic (quadratic) behavior that, irrespective of the electron density, occurs exactly at $\nu = 2$. This effect, termed {\it hidden symmetry} (HS) \cite{rashba}, results
from  a cancellation between the Coulomb interaction among the electrons in the conduction band and with the photo-generated hole in the valence band. It requires the square modulus of the electron and hole envelope functions to be similar in shape.
\par
Motivated by this scenario, here we report the magneto-PL study of QH states in 
coupled electron bilayers. For the bilayer with vanishing tunneling gap the intensity minima of the lowest energy emission line at 
$\nu _T =1$ and at $\nu _T =1/2$ represent unambiguous manifestations of the occurrence of
such inter-layer correlated quantum Hall states in magneto-PL.
The evolution of the magneto-PL line intensities in a bilayer with a finite value of the tunneling gap confirms the loss of pseudospin polarization at $\nu _T =1$ that arises
from excitonic correlations in the ground state and reveals
a similar but more pronounced effect at $\nu _T =2/3$. 
In addition, in both samples we observe the characteristic signature of the hidden-symmetry transition which, contrary to conventional single layer systems, is seen at 
$\nu_T =4$ due to the impact of the pseudospin degree of freedom. Indeed the HS requires both electrons and holes to be in the lowest LL. In double layers, because of the simultaneous presence of spin and pseudospin degrees of freedom,each LL consists of four sublevels with similar envelope function profile and hence the HS becomes valid for $\nu < 4$, independently of the presence of a finite tunneling gap.
\begin{figure}[!htp]
\centering
\includegraphics[width=0.35\textwidth]{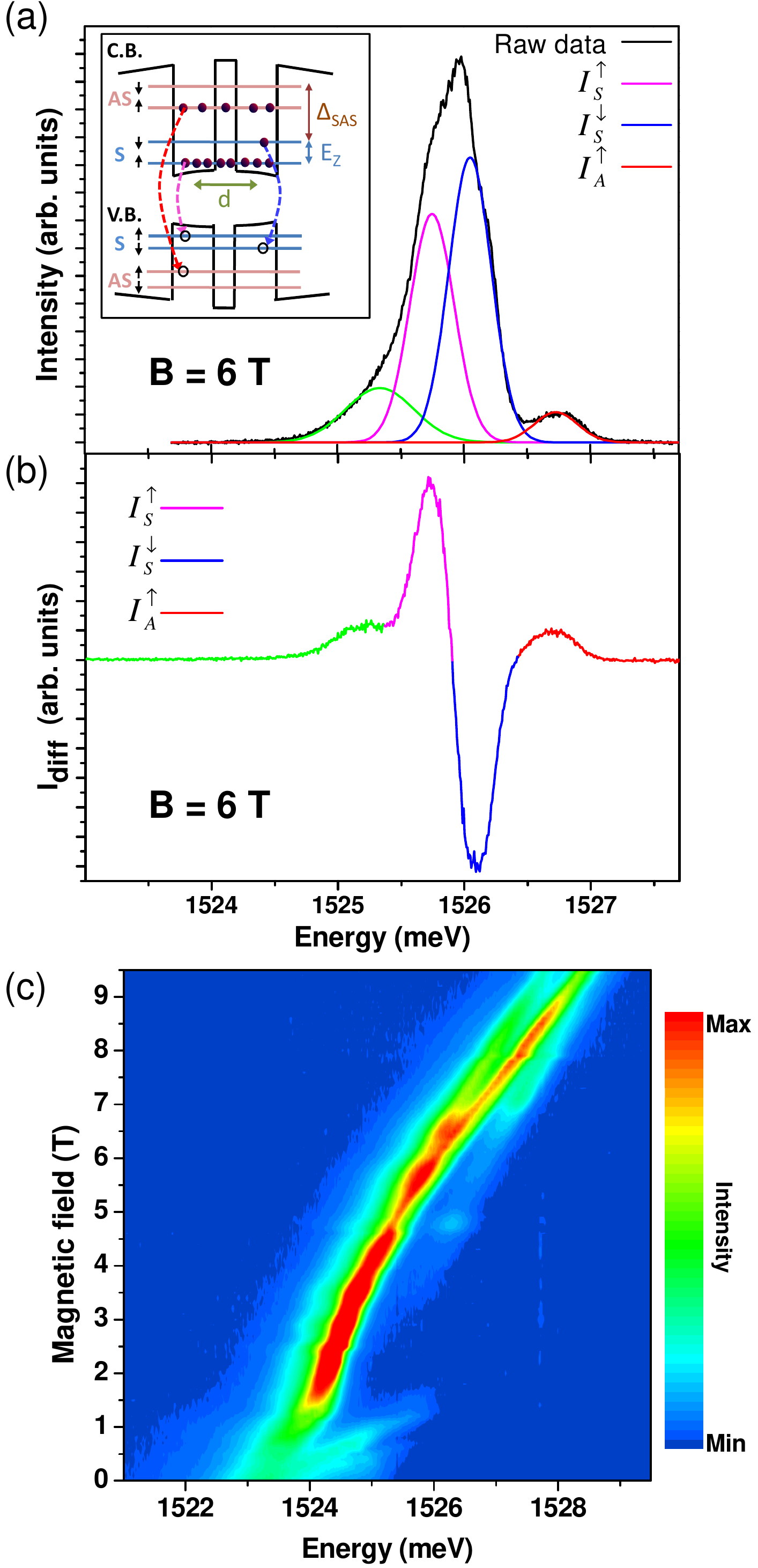}
\caption{(Color online) Data from the finite tunneling gap sample. (a) Representative left-circularly ($\sigma^-$) polarized
spectrum at B = 6 T and 50 mK fitted with Gaussian lines. The inset is a schematic representation
of the electron states in the lowest Landau level (LL) in the conduction
and valence bands. $\it S$ and $\it AS$ label the symmetric and anti-symmetric
combinations of the quantum-well levels, respectively. Each LL is further
split by the Zeeman term $E_Z$. (b) Difference of the $\sigma^+$ to the $\sigma^-$ spectra at 6 T.
(c) Color plot of the $\sigma^-$ polarized spectra in the range 0 - 9.5 T.}
\label{t_color}
\end{figure}
\par
Measurements were performed on samples mounted in a dilution
refrigerator with a base temperature of 50 mK
under light illumination. Two samples were studied. The first
is a nominally symmetric modulation-doped
$\textnormal{Al}_{0.1}\textnormal{Ga}_{0.9}\textnormal{As}/\textnormal{GaAs}$
double quantum well structure with AlAs barrier in between the wells,
having well width of 18 nm and barrier width of 7 nm. The large
barrier ensures that the tunneling gap is vanishingly small
($\Delta_{SAS} \rightarrow 0$). The total electron density is $n_T
\sim 6.9 \times 10^{10}$ $\textnormal{cm}^{-2}$ and the electron
mobility is above $10^6$ $\textnormal{cm}^2/\textnormal{Vs}$.
The other sample is also a nominally symmetric modulation-doped
$\textnormal{Al}_{0.1}\textnormal{Ga}_{0.9}\textnormal{As}/\textnormal{GaAs}$
double quantum well structure identical to the first one but with an $\textnormal{Al}_{0.1}\textnormal{Ga}_{0.9}\textnormal{As}$
barrier in between the wells, leading to a tunneling gap
at zero magnetic field of $\Delta_{SAS}$ = 0.36 meV \cite{Lu05}. This sample 
has a total electron density of $n_T \sim 1.1 \times 10^{11}$ $\textnormal{cm}^{-2}$
and electron mobility above $10^6$ $\textnormal{cm}^2/\textnormal{Vs}$.
\par
A perpendicular magnetic field was applied to the
electron bilayer. The magneto-PL
spectra were measured after excitation with a single-mode tunable
Ti-Sapphire laser at 795 nm. Laser power densities were
kept at $\sim 10^{-4}$ $\textnormal{W/cm}^2$ to avoid electron
heating effects and circularly-polarized configurations were
exploited to have access to spin states. A triple-grating spectrometer
equipped with a CCD detector was used to detect the emitted light.
\par
Figure \ref{t_color}(a) shows a representative left
circularly-polarized ($\sigma^-$) PL
spectrum at a magnetic field of 6 T from the sample
with a finite tunneling gap. The main PL lines (magenta and blue)
are assigned to recombination from the symmetric spin-up ($I_{S}^{\uparrow}$)
and spin-down ($I_{S}^{\downarrow}$) levels \cite{note1}, respectively,
while the higher energy peak (red line)
is linked to recombination of  electrons in the AS$\uparrow$ level ($I_{A}^{\uparrow}$).
This assignment is supported by the
circularly-polarized analysis reported in Fig. \ref{t_color}(b),
which shows that the magenta and red peaks are left circularly
polarized ($\sigma^-$), as opposed to the right circularly polarized
($\sigma^+$) blue peak.
Furthermore the energy separation of the blue and magenta lines
displays a linear dependence on the magnetic field as expected from Zeeman-split lines
with an effective Land\'e factor of $g_{eff} \approx 1.4$ (data non shown). 
The additional low-energy shoulder (green 
line in Fig. \ref{t_color}(a)) follows the evolution
of the main PL line. We ascribe it to a disorder assisted recombination and it will
not be further discussed in the following.
\par
\begin{figure}[!htp]
\centering
\includegraphics[width=0.35\textwidth]{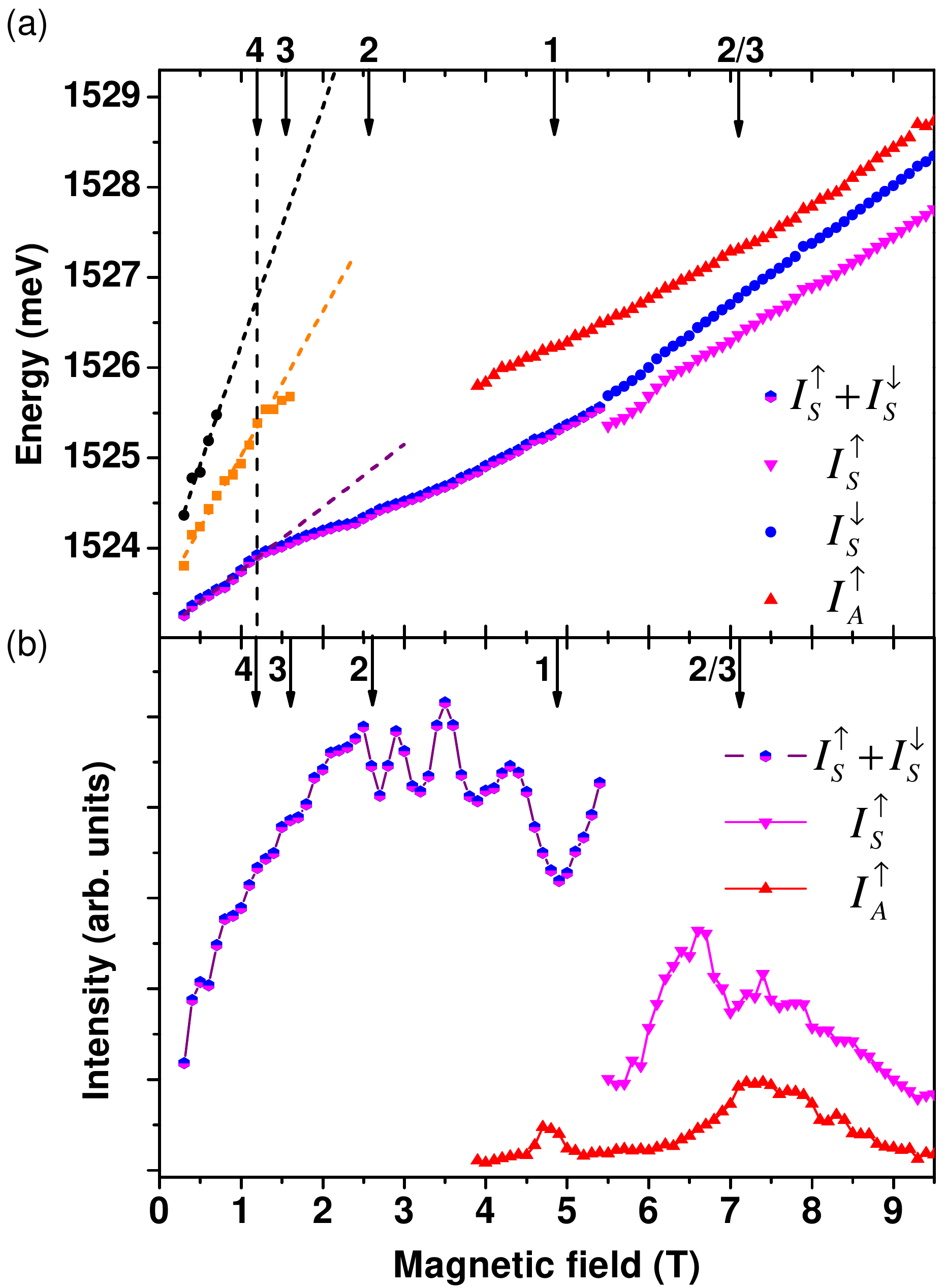}
\caption{(Color online) (a) Peak energies and (b) integrated intensities
vs magnetic field for the $\sigma^-$ polarized PL spectra from the
finite tunneling gap sample at 50 mK.
The filling factors of the QH states observed in transport experiments are indicated in
the upper axis. The short-dashed lines in (b) represent the best linear
fits to the energy data at low magnetic fields.}
\label{t_data}
\end{figure}

Figure \ref{t_color}(c) is a color plot of the magneto-PL in
$\sigma^-$ polarization. We can identify two different regions: a low-field
region (B $<$ 1.5 T) where a Landau fan of three peaks can be noticed,
and a high-field region (B $>$ 1.5 T) where the main emission line
deviates from the linear behavior and in addition it displays several
intensity oscillations. The plots of the peak energies and intensities,
are shown in Figs. \ref{t_data}(a),(b). 
\par
If (n,m) denotes the optical recombination of the electron in the $n$ LL
with the heavy-hole in the $m$ LL , then the linear
energy variations of the blue-magenta \cite{note2}, orange and black
peaks in Fig. \ref{t_data}(b) are
compatible with the (0,0), (0,2) and (1,1) recombinations, respectively \cite{note3}. 
\par
At $\nu_{T} = 4$ the magnetic-field dependence of the blue-magenta line energy
changes abruptly from linear to quadratic, indicating the formation of
a bound exciton. In analogy to magneto-PL studies in single layers
\cite{rashba} we interpret this changeover as due to the onset of the HS. This observation extends the validity of the HS to coupled bilayers 
where the lowest LL consists of four sublevels owing to the presence of both spin and pseudospin degrees of freedom.
\par
Figure \ref{t_data}(b) reveals several intensity oscillations
of the lowest energy emission line from the electrons in the lowest symmetric (pseudospin-up) LL. 
The minima at 2.6, 4.8 and 7.1 T are linked to the occurrence of QH states with
$\nu_T = 2$, $1$ and $2/3$, respectively, which are also observed in
transport measurements (data not shown here).
In addition, the emission from the antisymmetric (pseudospin-down) spin-up level displays 
maxima around $\nu_T = 1$ and $\nu_T = 2/3$,  suggesting that at these two QH states a fraction of electrons
populates the AS level as a consequence of a loss of pseudospin polarization. At $\nu_T = 1$
the loss of pseudospin polarization was previously observed in inelastic light scattering spectra
\cite{Lu05} and interpreted as a result of the formation of electron-hole excitonic pairs across $\Delta _{SAS}$.  
At $\nu_T = 2/3$ no evidence was reported so far.
\par
The emission intensity from the AS state increases
by a factor of two passing from $\nu_T = 1$ to $\nu_T = 2/3$, suggesting that
the loss of pseudospin polarization is more pronounced for the 2/3 state.
Indeed a simple estimate based on the ratio of the relative intensities of the S and AS emission lines and
on the $\nu_T = 1$ pseudospin polarization value of 36$\%$ reported previously \cite{Lu05} suggests that
for the 2/3 state the loss of pseudospin polarization is complete.
This result is in agreement with numerical studies \cite{num23}
that describe this state through the pseudospin unpolarized Halperin (3,3,0) wavefunction for
$d/l_B$ above some critical value depending on $\Delta_{SAS}$. In fact, for sufficiently large
$d/l_B$ the energetic advantage of localizing electrons in opposite layers (as in the (3,3,0) state) outweighs the tunneling energy cost.
\par
\begin{figure}[!htp]
\centering
\includegraphics[width=0.35\textwidth]{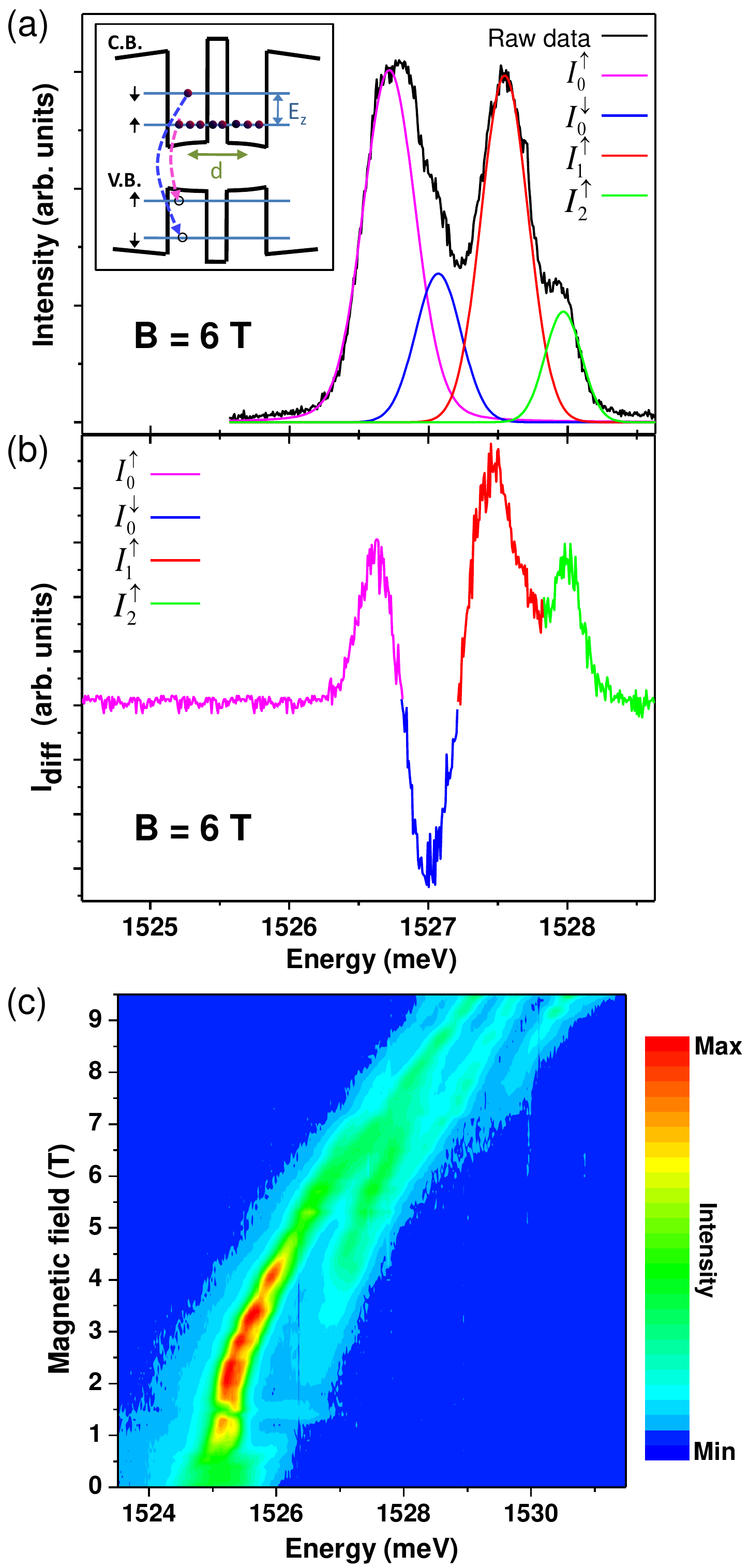}
\caption{(Color online) PL data from the sample with
vanishing tunneling gap at 50 mK. (a) Representative spectrum in $\sigma^-$
polarization at B = 6 T fitted with Gaussian lines. The inset represents a
schematic of the spin-split states in the lowest Landau level in the
conduction and valence bands. Each spin state has a double pseudospin
degeneracy. (b) Spectrum resulting from the difference between the $\sigma^-$
and the $\sigma^+$ polarized emissions at 6 T.
(c) Color plot of the $\sigma^-$ polarized spectra in the range 0 - 9.5 T.}
\label{not_color}
\end{figure}
We focus now on the sample with vanishing tunneling gap. A representative
$\sigma^-$ polarized emission spectrum from this sample at B = 6 T is shown
in Fig. \ref{not_color}(a). Four emission lines are identified, which we
label $I_0^{\uparrow}$, $I_0^{\downarrow}$, $I_1^{\uparrow}$, and $I_2^{\uparrow}$,
in increasing order of energy. The polarization analysis shown in Fig.
\ref{not_color}(b) indicates that the $I_0^{\uparrow}$, $I_1^{\uparrow}$, and
$I_2^{\uparrow}$
lines are $\sigma^-$ polarized and we link them to the recombinations
of electrons in the lowest spin-up LL (see inset to Fig. \ref{not_color}(a)) with different heavy-hole levels.
The $I_0^{\downarrow}$, on the contrary, involves the recombination of spin-down electrons.
The magnetic field variation of the energy difference
between $I_0^{\uparrow}$ and $I_0^{\downarrow}$ is linear as expected for 
Zeeman-split lines and yields an effective Land\'e factor of $g_{eff} \approx 1.2$,
similar to the value obtained for the other sample.
The magnetic field evolution of the $\sigma^-$ PL emission is reported in Fig. \ref{not_color}(b). 
\par
\begin{figure}[!htp]
\centering
\includegraphics[width=0.35\textwidth]{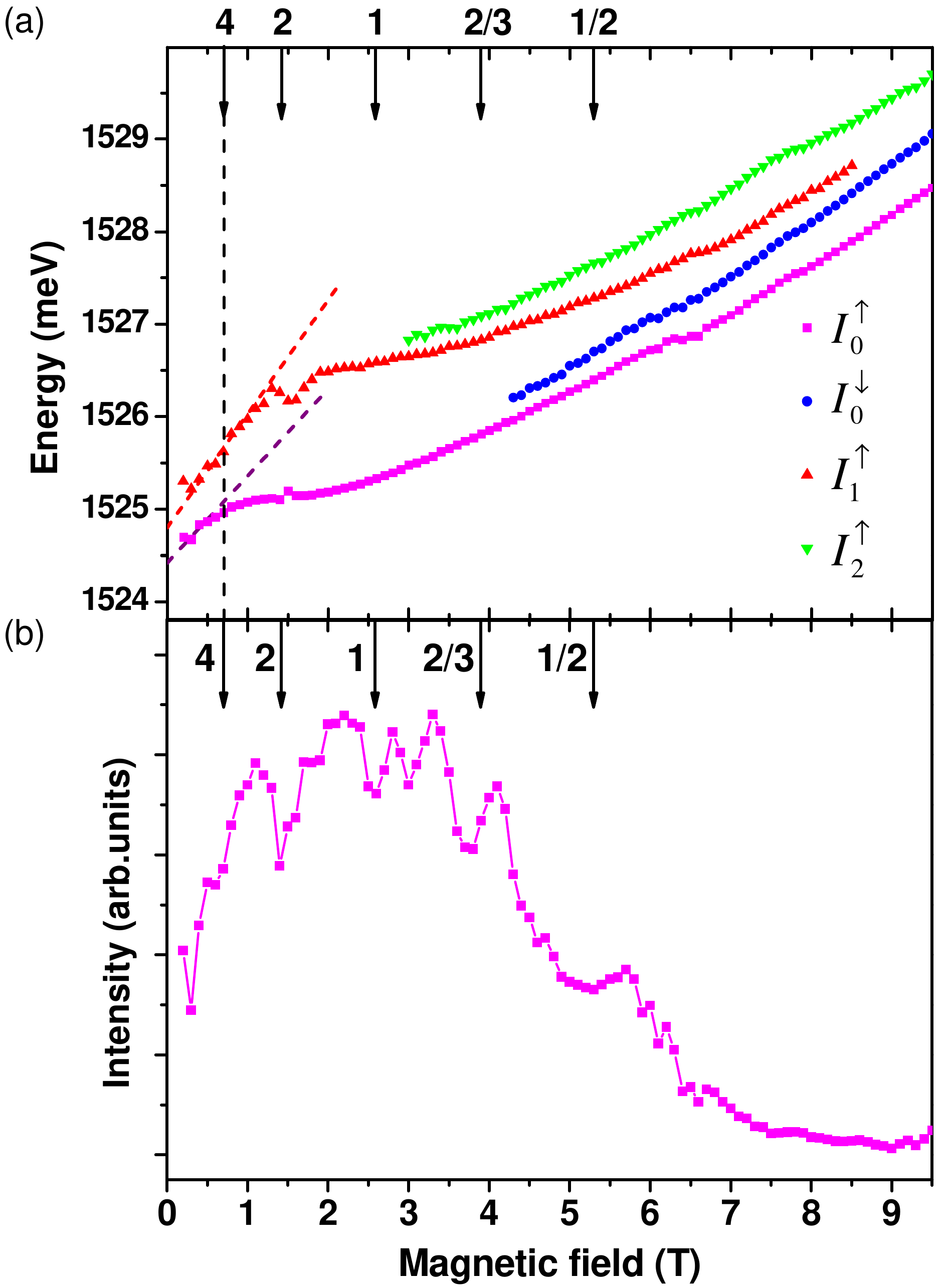}
\caption{(Color online) (a) Peak energies and (b) $I_0^{\uparrow}$ integrated
intensity as a function of magnetic field for the $\sigma^-$ polarized PL
spectra from the zero tunneling gap sample at 50 mK. The filling factors of
the observed QH states are shown in the upper axis. Short-dashed lines in
(a) represent the best linear fits to the energy curves at low magnetic fields.}
\label{not_data}
\end{figure}
The peak energies for all four lines are shown
in Fig. \ref{not_data}(a). At low magnetic fields, the PL energies vary
linearly with B. Using the same values as above for the electron and heavy-hole
effective masses, we attribute the $I_0^{\uparrow}$, and $I_1^{\uparrow}$ emissions to 
the (0,0), (0,1) recombinations, respectively. Again the lowest
energy line ($I_0^{\uparrow}$) displays an abrupt change-over from linear
(single particle) to quadratic (excitonic) behavior at $\nu_T = 4$, suggesting
the impact of the HS and of the pseudospin degree of freedom also in this case. 
The linear-to-quadratic change occurs at $\nu_T = 2$ for the $I_1^{\uparrow}$ line.
Indeed this emission line involves holes from a higher LL ($m = 1$) and therefore it
is not subject to the HS mechanism.
The different behavior between the lowest and the
higher energy lines was also observed in single layers \cite{yoon}.
\par
The magnetic field positions of the QH states at $\nu_T$ = 4, 2, 1, 2/3 and 1/2 as identified in magneto-transport data (not shown) 
are indicated with arrows in Fig. \ref{not_data}. The lowest energy line $I_0^{\uparrow}$ displays intensity minima
(see Fig. \ref{not_data}(b)) in correspondence to the occurrence of such QH states.
The intensity minima appear independently from the value of the laser
excitation wavelength (data not shown), which rules out the possibility that they could result
from magnetic field-induced changes in the absorption. 
The quenching of the emission is indeed a manifestation of the QH states and can be linked to the reduction of the optical matrix element
associated with the onset of QH phases. The latter follows from the localization
of electrons and holes in the disorder potential, which increases in the gapped QH phases
because of the reduced electron screening  \cite{gold91}.
We remark that the observed QH states with $\nu_T = 1$ and 1/2 are genuinely
linked to the impact of inter-layer correlations. In particular the $\nu_T = 1/2$
state has no counterpart in single-layer single-component systems.
\par
Finally the magneto-PL energies vary smoothly with B for $\nu_T < 4$ in both samples (see Figs. \ref{t_data}(a) and \ref{not_data}(a))
and do not provide evidence of QH states. This behavior is a consequence of the HS. 
\par
In conclusion we have studied the magneto-PL spectra in coupled bilayers in the QH regime. 
The evolution of the intensities of the emission lines in a magnetic field
reveals a loss of pseudospin polarization at $\nu_T = 1$ and 2/3 in a sample
with a finite moderate value of $\Delta_{SAS}$ and signals the occurrence of inter-layer
correlated QH states at $\nu_T = 1$ and 1/2 in the vanishing $\Delta_{SAS}$ sample. The energy evolution of the emission lines
reveals the role of the hidden symmetry at magnetic fields above $\nu_T = 4$.
Magneto-PL appears as a promising technique to investigate the role of inter-layer correlation
in bilayers. Future experiments in slightly asymmetric double layers can finely probe the role of inter-layer electron interactions
at $\nu_T$ = 1 or in the fractional QH regime.
\par
We thank I. Bar-Joseph for stimulating discussions.
This work is supported by ITN project INDEX. A. Pinczuk was
supported by NSF under Grant No. DMR-0803445 and by the Nanoscale
Science and Engineering Initiative of NSF under award No.CHE-0641523. The work at Princeton was partially funded by the Gordon and Betty Moore Foundation as well as the National Science Foundation MRSEC Program through the Princeton Center for Complex Materials (DMR-0819860). 

\nocite{*}

\begin{thebibliography}{40}
\bibitem{note1} The emission from the spin-down level should only appear in
the $\sigma^+$ polarized spectra. However,  heavy hole-light hole
mixing effects in the valence band lead to a partial circular polarization
of all emission peaks, which are thus observed in both circular polarizations in both bilayer samples studied here.
\bibitem{note2} The blue-magenta (or $I_S^{\uparrow} + I_S^{\downarrow}$) peak
represents the overall emission from the symmetric level. In fact, for
B $<$ 5.5 T the Zeeman energy is smaller than the emission linewidth and
 the recombinations from the S$\uparrow$ and S$\downarrow$ levels
are not resolved.
\bibitem{note3} In this analysis
we have assumed effective in-plane masses of
$m_e = 0.068 m_0$ for the electron and $m_{hh} = 0.38 m_0$ for the
heavy-hole, where $m_0$ is the electron mass in vacuum.
\end{thebibliography}

\end{document}